\documentclass[journal,electronic]{vgtc}                % final (journal style)
\ifpdf%                                % if we use pdflatex
  \pdfoutput=1\relax                   % create PDFs from pdfLaTeX
  \usepackage{graphicx}                % allow us to embed graphics files
  \DeclareGraphicsExtensions{.pdf,.png,.jpg,.jpeg} % for pdflatex we expect .pdf, .png, or .jpg files
\else%                                 % else we use pure latex
  \usepackage{graphicx}                % allow us to embed graphics files
  \DeclareGraphicsExtensions{.eps}     % for pure latex we expect eps files
\fi%

\graphicspath{{figures/}{pictures/}{images/}{./}} % where to search for the images

\usepackage{microtype}                 % use micro-typography (slightly more compact, better to read)
\PassOptionsToPackage{warn}{textcomp}  % to address font issues with \textrightarrow
\usepackage{textcomp}                  % use better special symbols
\usepackage{mathptmx}                  % use matching math font
\usepackage{times}                     % we use Times as the main font
\usepackage{cite}                      % needed to automatically sort the references
\usepackage{tabu}                      % only used for the table example
\usepackage{booktabs}                  % only used for the
\usepackage{float}
\usepackage{todonotes}

\usepackage{scalerel}

\usepackage{xspace}
\newcommand{\ie}{{i.e.,}\xspace}
\newcommand{\eg}{{e.g.,}\xspace}

\newcommand{\ea}{{et~al.}\xspace}

\newcommand{\etc}{{etc.}\xspace}

\onlineid{0}

\vgtccategory{Research}
\vgtcpapertype{please specify}

\title{Beyond Heuristics: Learning Visualization Design}
\author{Bahador Saket\thanks{\vspace{-0.8em}Bahador Saket and Dominik Moritz contributed equally to this work.}, Dominik Moritz$^{*}$, Halden Lin, Victor Dibia, {\c{C}a}{\u{g}}atay Demiralp, Jeffrey Heer}
\authorfooter{
\item
 Bahador Saket is with Georgia Tech. E-mail: saket@gatech.edu.
\item
 Dominik Moritz, Halden Lin, and Jeffrey Heer are with University of Washington. E-mail: {domoritz, haldenl,jheer}@uw.edu.
\item
 \c{C}a\u{g}atay Demiralp is with Columbia University. E-mail: cd3057@columbia.edu.
 \item
 Victor Dibia is with IBM Research. E-mail: dibiavc@us.ibm.com.
}

\shortauthortitle{Saket and Moritz \MakeLowercase{\textit{et al.}}: Beyond Heuristics: Learning Visualization Design}
\abstract{
  In this paper, we describe a research agenda for deriving design 
  principles directly from data. We argue that it is time to go beyond manually curated and applied visualization design guidelines. We propose learning models of visualization design from data collected using graphical perception studies and build tools powered by the learned models. To achieve this vision, we need to 1) develop scalable methods for collecting training data, 2) collect different forms of training data, 3) advance interpretability of machine learning models, and 4) develop adaptive models that evolve as more data becomes available.
}


\CCScatlist{ % not used in journal version
 \CCScat{K.6.1}{Management of Computing and Information Systems}%
{Project and People Management}{Life Cycle};
 \CCScat{K.7.m}{The Computing Profession}{Miscellaneous}{Ethics}
}

\vgtcinsertpkg

\begin{document}

\firstsection{Introduction}\label{sec:intro}

\maketitle

The demand for data visualization has significantly grown in recent years with the increasing availability and complexity of digitized data across everyday domains. By visually encoding data properties, visualizations aim to enhance data understanding by leveraging human visual perception, which has evolved for fast pattern detection and recognition. Understanding the effectiveness of a given visualization in achieving this goal is a fundamental pursuit in data visualization research and has important implications in practice. Visualization researchers regularly conduct empirical studies to investigate how people decode and interpret visualizations (e.g.,~\cite{cleveland1984,kim2018,talbot:infovis11,saket2018,Szafir2018}).
% simkin:jasa87, heer:chi09, kong:infovis10, Szafir2018, cleveland1985graphical, cleveland1984graphical
Such empirical studies are important for understanding and improving the
effectiveness of visualizations. Indeed, design guidelines and heuristics that we use today in data visualization are an accumulation of what we have learned from such empirical studies over decades. 

It is not, however, always possible to distill guidelines through analysis of
data collected by empirical studies due to various confounding factors. Even
when this is possible, derived guidelines might be inadequate for conveying the
subtleties present in user data or might be marred by those providing the
guidelines. Moreover, design guidelines provided by empirical studies often make
their way to visualization tools slowly for two main reasons. First, our design
knowledge is continually evolving as visualization researchers regularly
publish results from empirical studies and provide new sets of guidelines
for designing effective visualizations. Second, designers of visualization
tools need to spend a significant amount of time and effort to manually
incorporate these guidelines.  
  
In recent years, there has been an increasing trend to publish data
collected from empirical studies with increased awareness of the importance of
replicability and reproducibility. Researchers have made large datasets of
experimental data of visualizations' effectiveness publicly available (e.g.,
~\cite{heer:chi09,Szafir2018,kim2018,saket2018}).
% kong:infovis10,Harrison2014,2016-beyond-webers-law
We advocate learning models from this data and building tools that automatically apply the learned model. We believe machine learning models provide a practical
opportunity to implicitly engineer the insights provided by empirical user
performance data into visualization systems. The main advantage is that new
guidelines can be applied in practice faster in an unbiased and reproducible
manner.

In this paper, we discuss how we might automatically derive visualization
design principles from experimental data in a scalable way by using machine
learning. We further categorize and discuss the existing approaches for learning visualization designs to illustrate feasibility and how future systems fit into these categories. We argue that visualization design principles used today are also
derived from data. However, these principles had to be abstracted by a
visualization researcher, taught by a teacher, and applied manually by the
designer. We propose that the next step for the research community is to curate
a knowledge base of design principles that can be applied automatically. 
Next, we should aim to learn from data both design principles and how to trade off among them.  As more data becomes available, we may one day be able to learn
visualization design end-to-end. In the following, we describe each of these
steps, and future research directions to achieve this vision.

\section{Visualization Recommendation Systems}\label{sec:related}

In this section, we discuss existing and previous work on automated visualization design and recommendation engines that incorporate guidelines derived from graphical perception studies. We categorize these systems into \textbf{knowledge-based}, \textbf{data-driven}, and \textbf{hybrid} visualization design tools. These categories are used in recommender systems~\cite{aggarwal2016recommender} and represent where the knowledge that the system uses to recommend visualizations comes from. \autoref{tab:systems} gives an overview of existing machine learning and knowledge-based systems.

% \dom{Talk about this as recommender systems: knowledge based system (APT, Compass, ...), data driven recommender systems (Kopol, Kim et al, ...), hybrid systems (Draco)}

% % In this section, we first explain how designers of many of the existing
% % automated visualization tools manually incorporate guidelines derived from
% % empirical studies to improve visualization pruning and ranking in these tools.
% % We then discuss some of the recent studies that took the initial steps to go
% % beyond manual incorporation of design guidelines by using machine learning
% % models instead of manual incorporation of these guidelines.      

% In this section, we first discuss prior work on automated visualization
% design that manually incorporate guidelines derived from graphical perception
% studies. We then introduce recent systems and studies that took the
% initial steps to use machine learning to learn design guidelines.

\subsection{Knowledge-Based Systems}

A large body of existing automated visualization design tools focuses on
suggesting visualizations based on user-defined constraints (such as which data
attributes to visualize) and design constraints. Mackinlay's
APT~\cite{Mackinlay1986} leverages a compositional algebra to enumerate the space
of visualizations. It then uses \emph{expressiveness} and \emph{effectiveness}
criteria based on the work by Bertin~\cite{bertin1983semiology} and Cleveland
\& McGill~\cite{cleveland1984} to prune and rank visualizations.
\emph{Expressiveness} refers to the ability of a visualization to convey all and only
the facts in the data. \emph{Effectiveness} refers to the ability of a
visualization when the information it conveys is more readily perceived than
with other visualizations. The SAGE project~\cite{Mittal1998} extends APT by
taking into account additional data properties such as cardinality, uniqueness,
and functional dependencies. Tableau's Show Me~\cite{Mackinlay2007} introduces
a set of heuristic rules to recommend visualizations. Following this line of
work, CompassQL~\cite{Wongsuphasawat2016} also uses similar heuristic rules to develop expressiveness
and effectiveness criteria to evaluate visualization options. However,
CompassQL extends the earlier work by using a set of hand-tuned
scores to specify criteria such as space efficiency and legibility based on
visualization and data properties.

Many of the automated design tools discussed above prune and rank candidate visualizations based on a set of manually curated design guidelines derived from previous empirical studies. Designers of these tools often spent a considerable amount of time and effort to incorporate the findings of the previous and
current empirical studies. 

% \subsection{Data Driven Models for Visualization Design}

% Earlier work developed low-level perceptual models such as Stevens'
% Weber-Fechner Law~\cite{fechner1966elements}, Stevens' Power
% Law~\cite{stevens1957psychophysical}, and CIELAB color space that are based on
% the data collected from empirical studies. In a more recent work, Demiralp et
% al.~\cite{perceptualKernels} also used the data collected from an empirical
% study to created a model of perceptual kernels.

\begin{table*}[htb]
	\centering
	\begin{tabular}{l l l l l l l}
	% \toprule
	System & Recommender Type & Modeling Approach & Learning & Model & Input Features & Design Space \\
	\midrule
	\texttt{VizDeck}    & Data Driven       & Basic Features        & Yes   & Linear               & 5 Data properties per field        & 9+ Types \\
	\texttt{Kopol}      & Data Driven       & Basic Features        & No    & Tree                 & Data properties, Task              & 5 Types \\
	\texttt{Kim et al.} & Data Driven       & Basic Features        & No    & Linear               & Data properties, Task              & 12 Scatterplots \\
	\texttt{Data2Vis}   & Data Driven       & Raw Features          & Yes   & Deep                 & Raw Dataset                        & Vega-Lite \\
	\texttt{APT}        & Knowledge Based   & Hand Tuned            & No    & Rules                & Data Types, Importance ordering    & APT \\
	\texttt{CompassQL}  & Knowledge Based   & Hand Tuned            & No    & Linear               & Partial Specification              & Vega-Lite \\
	\texttt{Draco}      & Hybrid            & Learning Trade-Offs   & Yes   & Linear               & Partial Specification, Task        & Vega-Lite \\
	% \bottomrule
	\end{tabular}
	\vspace{0.5em}
	\caption{Comparison of different approaches to model automated visualization design by the type of recommender(\autoref{sec:related}), the modeling approach based on the kind of features that are used (\autoref{sec:models}), whether the approach uses machine learning with genralization, the type of model, the input to the model, and the design space.\label{tab:systems}}
	\vspace{-1em}
\end{table*}

\subsection{Data-Driven Systems}

The visualization literature is no stranger to data-driven models elicited through human-subject experiments.  For example, 
low-level perceptual models
such as Weber-Fechner Law~\cite{fechner1966elements}, Stevens' Power
Law~\cite{stevens1957psychophysical},
% color-name distance~\cite{heer:chi12}
and perceptual kernels~\cite{perceptualKernels}
are all based on fitting parametric and
non-parametric models to empirical user data, informing low-level visual
encoding design. While data-driven models are prevalent, using
data-driven models for automated visualization design is a nascent area
and only a handful of papers exist to date.

% As machine learning libraries (e.g., scikit-learn, TensorFlow) and platforms
% (e.g., AutoML, Auto-WEKA), and the data collected from empirical studies are
% getting more and more accessible, there is a great opportunity to learn models
% from this data and build tools that apply the learned models.
% \autoref{tab:systems} summarizes different systems for learning visualization
% design. \\

With advances in machine learning, more researchers in the visualization community started taking initial steps towards developing machine learning models to recommend visualizations. A machine learning model tries to best predict what visualization
is most appropriate based on the given inputs (e.g., tasks, data attributes, etc.). Developers of visualization tools need to hand-craft informative, discriminating, and independent features for learning such models. To the best of our knowledge, VizDeck~\cite{vizdeck2012} was the first attempt at learning a recommendation model for high-level visual design. VizDeck is a web-based visualization recommender tool designed for exploratory data analysis of unorganized data. Using users' up and down votes on a gallery of visualization, VizDeck learns to recommend charts that the user is most likely going to vote up. It does so by learning a linear scoring function over statistical properties of the dataset. Visualizations are picked from a corpus of possible candidate visualizations.

Saket~\ea~\cite{saket2018} recently conducted a crowdsourced study to evaluate the effectiveness of five visualization types (Table, Line Chart, Bar Chart, Scatterplot, and Pie Chart) across 10 different visual analysis tasks and from two different datasets. Based on their findings, they developed Kopol\footnote{\url{https://kopoljs.github.io/}}, a mini JavaScript prototype visualization recommender that uses a decision tree to suggest visualizations based on the given tasks and data attribute types. Kim~\ea~\cite{kim2018} also recently developed a model for 12 scatterplot encodings, the task type, and the cardinality and entropy of some data fields. During their crowd-sourced experiment, Kim~\ea had users perform tasks for a combination of features and used results to create a ranking.

Luo~\ea~\cite{deepEye:2018} conducted a study in which 100 participants annotated 33,412 visualizations as good/bad, and provided 285,236 pairwise comparisons between visualizations. They used 42 public datasets to create the visualizations for their experiment. Luo~\ea then developed DeepEye~\cite{deepEye:2018}, a visualization recommender system that uses a binary classifier to decide if a visualization is good or bad, and a supervised learning-to-rank model to rank the visualizations based on users' preferences data collected in their experiment. 

% VizRank is a different system which is applied on classified data to automatically select the most useful data projections. Given a data set and a visualization type, VizRank prunes possible projection methods and provides the data analyst with a ranked list of projection methods along with the assessment of their effectiveness.

A more recent trend in machine learning is to learn models on all available features instead of hand-crafting good features---a labor intensive and often biased process. For example, Data2Vis~\cite{data2vis} is a neural translation model that generates visualization specifications in the Vega-Lite grammar from descriptions of a dataset. The model was trained on thousands of example pairs of data and visualizations recommended by CompassQL.

%  Unlike Draco, Data2Vis does not work in a structured domain and so the model
% has to learn the Vega-Lite grammar in addition to the recommendation model. 

\subsection{Hybrid Systems}

In knowledge-based systems, the system designer provides the knowledge about visualization design. In data-driven systems the system learns a recommendation model from data. Hybrid recommender systems are both knowledge-based and data-driven. The system designer has full control over the recommendation process but can augment the knowledge base with machine learning. Recently, machine learning experts argued that learning with structure must be a top priority for AI to achieve human-like abilities~\cite{relationalmodels}. 

Draco~\cite{draco} is a formal model that represents (1) visualizations as sets of logical facts and (2) design principles as a collection of hard and soft constraints over these facts. Using constraints, we can take theoretical design knowledge and express it in a concrete, extensible, and testable form.
Compared to VizDeck and Kopol, Draco's visualization model covers a larger set of visualizations supported by Vega-Lite~\cite{Satyanarayan2017}.
Similar to CompassQL~\cite{Wongsuphasawat2016}, Draco's default model uses rules that are informed by empirical studies but formalized by experts. To avoid ad-hoc choices and support an evolving knowledge base, Draco can learn how to trade off among competing design rules using a simple linear model. Draco's learning algorithm uses learning to rank, a machine learning method that enables it to learn a preference model from ordered pairs of complete visualizations. Using this method, Draco can learn without the need to normalize scores between perceptual studies with different methods and conditions.

\section{Features for Modeling Visualization Design}\label{sec:models}

% \hl{[Alternative classification] In this section, we discuss the different approaches to automated visualization design with respect to the features that they use. These approaches can be roughly classified into three groups. The first is a top-down approach to encoding visualizations as features. Defining features as visualization type falls under this category, wherein a collection of elements is grouped under an umbrella term (e.g. scatter plot, bar chart, etc.) that describes the graphic as a whole. From here, more nuanced categories can be defined, splitting the visualization space into smaller pieces. The next is a bottom-up approach to encoding visualizations. Here, features are defined by using an underlying grammar of graphics, which defines the visualization space. This allows for more sophisticated features to be engineered up from the underlying grammar, with design principles being encoded in these engineered features. Finally, the third approach is also a bottom-up approach, using a grammar of graphics as a basis, but to forgo human input and design principles in lieu of automatically learning features using the model itself. Table 1 provides an overview of the systems discussed here.}

In this section, we discuss the different approaches to automated visualization design with respect to the features that they use. In particular, we discuss building models from basic features (visualization type, data properties, task), feature engineering, learning design rules, and learning without feature engineering. \autoref{tab:systems} provides an overview over the systems discussed here.

In the discussion below, we assume that a machine learning model for visualization design takes as input a set of features that can include some specification of the visualization, data, and task and outputs a corresponding score. This score can represent how preferred a visualization is (based on effectiveness, how easy it to read the visualization, \etc). The magnitude of the score may not be meaningful, but it provides a rank ordering of the feature vectors (\ie possible designs). An automated visualization design system can use any model of this form by enumerating possible designs and recommending the one with the highest score.

% A machine learning model is the output generated when we train our machine
% learning algorithm (e.g., decision tree, SVM, \etc) with our training dataset
% (e.g., datasets collected from empirical studies).
% % We encourage the visualization community to create models using the available
% % datasets rather than creating and applying design guidelines manually.
% Building a supervised machine learning model requires two steps. First, a
% domain expert engineers a set of features. A feature is an measurable property
% or characteristic of a phenomenon being observed (e.g., type of a
% visualization, data attribute type, etc.). Selecting informative,
% discriminating, and independent features is a crucial step for learning
% accurate models. Then, a machine learning expert needs to select a suitable
% algorithm that uses data to train a model that predicts an output based on the
% input that is an assignment of values to the set of features.

% \begin{figure}[H]
%  \centering
%   \includegraphics[width=.8\linewidth]{pictures/learning}
%   \caption{A machine learning model can lean end-to-end from user input to visualizations. An alternative to reduce the amount of data needed for learning and add interpretability, the features (design rules) can be learned independently from the trade-offs among rules.}~\label{fig:learning}
% \end{figure}

\subsection{Modeling Using Basic Features}
\label{sec:basic}
The simplest and fastest way to model a score is to use the results of experimental studies of effectiveness. These models use data such as the type of visualization, data properties, and task type as features. For example, Kopol by Saket~\ea~\cite{saket2018} and Kim~\ea~\cite{kim2018} are examples of this approach. They use data such as tasks, visualization types, and data attribute types as features. VizDeck~\cite{vizdeck2012} is another example of a system that uses basic features. Instead of learning from effectiveness studies, VizDeck~\cite{vizdeck2012} uses users' up and down votes to change the scoring of each visualization alternative. In addition to the visualization type, VizDeck's model uses statistics of the data as input.

% Since the design space in these two papers is small, the model can have training data for each possible combination of input features. The two models do not need to generalize beyond these examples. While the rankings in the research paper can be used to look up the most effective encoding for each combination of data properties and task, these models miss generalization (generalize beyond the examples in the training set) as a core aspect of machine learning models~\cite{useful}. 

While it is often simple and fast to use the results of experimental studies to learn models (similar to Kopol~\cite{saket2018} and Kim~\ea~\cite{kim2018}), the design space of these models is small since they only support a small set of visualization types and a basic set of features. In practice, visualization designers might want to consider a larger design space. Going forward, the visualization community needs to develop models that cover a much broader design space. To effectively discriminate data in a larger design space, models need to use more expressive features. In the next section, we discuss a method for learning generalizable models with expressive features for visualization design.
% In summary, these models are limited in what designs they can express. For example, VizDeck and Kopol only support a handful of visualization types, which are umbrella terms that group collections of designs that designers might consider. The design space is small because models in this sections use a small set of basic features. However, in practice, a designer might consider a much larger design space. Going forward, the visualization community needs to develop models that cover a much broader design space. To effectively discriminate data in a larger design space, models need to use more expressive features. \bs{I still think we need a transition here !!! suddenly jumping to feature engineering feels odd!}
% Feature engineering is the process of creating and deciding on a set of features that the machine learning model can use to effectively evaluate the design space. It is considered the most important aspect of any machine learning project~\cite{useful}. In the next section we discuss a scalable method for feature engineering for visualization design.

\subsection{Using Visualization Design Rules as Features}
Another method for automating visualization design is to use the violations of design rules as features for learning models. A design rule is a predicate over properties of a visualization, the user's task, and the data. For example, ``Do not use the same field for the x and the y channel of a scatterplot'' describes the properties scatterplot and the same field on x and y should not occur together. Assuming a set of design rules, the feature vector representation of a visualization design is the number of times a design rule is being violated by the design. For example, a design that does not violate any design rules can be represented as a feature vector with only zeros. We can use these feature vectors to learn a model. Since each feature corresponds to a design rule, the weight of each feature in the learned model is a measure of the relevance of its corresponding rule. Many of the design rules that we use today are not always prescriptive. Today's visualization designers need to decide what rules they prefer for the specific visualization they are working on. Machine learning models use statistical models to handle uncertainty and noise in training data and are thus well equipped to handle design rules that are not prescriptive. A model that uses design rules as features learns the trade-offs among competing rules from data and is more deterministic than human designers when applying them.

Recently, Moritz~\ea proposed Draco~\cite{draco}, a method to derive feature vectors for a machine learning model from Vega-Lite specifications~\cite{Satyanarayan2017}. This allows for more sophisticated features to be composed, through feature-engineering, from the underlying grammar. The main idea in Draco's learning approach is to use the violations of \emph{design rules as features} as outlined above.

% Vega-Lite is a grammar that describes a combinatorial visualization space of many possible designs. 

The main advantage of this approach over pure data-driven learning is that the model can incorporate existing knowledge into the algorithm and thus learn a model that generalizes from little data. This approach is also a generalization of learning from basic features (\autoref{sec:basic}), as every feature about the visualization type or the data can be written as a predicate over the structural representation of the visualization (\eg in Vega-Lite) or the data schema. For example, to support tasks, we may just add rules for each task. However, one of the main challenges with this approach is the limited availability of design rules encoded in the system. Researchers should work on systematically enumerating design rules and encode them in a machine readable form so that they can be used as features in machine learning systems. In the next section, we discuss how machine learning may support them in this endeavour.

% Over the last century, the visualization community has accumulated many design rules. Insight from research such as the work of Bertin and Cleveland \& McGill already form the basis of automated visualization design systems. Instead of learning all these rules again (implicitly encoded in the complex weights of a machine learning model), learning based on rules lets us build automated design systems that use existing knowledge and models learned from data. For example, strict rules can encode the grammar of a visualization language and exclude invalid visualization specifications. This alleviates the need for providing the model with invalid examples.

% This approach is a generalization of learning from basic features as every feature about the visualization type or the data can be written as a predicate over the structural representation of the visualization (\eg in Vega-Lite) or the data schema. For example, to support tasks, we may just add rules for each task.

% Ignoring the availability of training data to learn the trade-offs between competing rules, the main concern with this approach is the availability of design rules encoded in the system. Researchers should work on systematically enumerating design rules and encode them in a machine readable form so that they can be used as features in machine learning systems. In the next section, we discuss how machine learning may support them in this endeavour.

\subsection{Learning Features from Data}
%\bs{after reading 3.3 many times, I am not sure if I can truly understand this. IT is using so many technical words (e.g., Markov logic models, Bayesian reasoning, logic programming) without really describing them to the readers. I understand that these stuff might be very clear to you guys, but we should consider that readers of this paper range from technical CS people to those with psychology background. I suggest either clarifying/simplifying this section or removing it.}

Systems that use design rules for feature engineering and then learn a linear
scoring function over a vector of violations are limited by these rules.
Design rules relate properties of a
visualization, and learning these relations is known in machine
learning as structure learning. For example, inductive logic programming methods can
infer logic programs from specific instances~\cite{Quinlan1990}. However, many
algorithms assume no or very little noise. Design rules, however, are not prescriptive and have exceptions. For example, the design rule that quantitative scales should start at zero does not make sense when the data has no meaningful zero such as temperature data. However, that does not mean that the zero rule should not be used.
Markov logic networks~\cite{Richardson2006} are statistical models that support Bayesian reasoning and thus can handle this uncertainty well. Their structure can be learned~\cite{Kok2005}.
% ILASP~\cite{Law2015} is a logic-based learning system that can learn preferences in answer set programs and could be used to learn rules for Draco.
It remains an open question whether these learning methods produce reasonable design rules from the experimental data that is available today.
Moreover, more work is needed to investigate how to design an experiment to collect data that results
in reliable rules. One approach may be to learn rules from data but have
experts confirm them.

The key to learning structured rules is the availability of large high quality datasets of examples.
However, as more data becomes available, we may be able to skip the feature engineering step and
learn an end-to-end model, as we discuss in the next section.

\subsection{Learning without Feature Engineering}

A recent trend in machine learning is to learn models on all available
features instead of hand-crafting good features. In particular, deep learning models shift the burden of feature engineering to the system by automatically learning abstract features and  complex relationships that would otherwise be difficult to capture by humans. This also means models can be more flexible. Machine learning on the full data has led to impressive results in computer vision and machine translation among other areas. However, deep learning models in particular are extremely data hungry
often requiring millions of training examples. The data also needs to cover a
large fraction of the design space. For example, to learn a visualization
design system that synthesizes visualizations not just from templates but a
grammar, the model needs to first learn the grammar. For instance, Data2Vis~\cite{data2vis} uses $215\,000$ training examples to learn a subset of
the Vega-Lite grammar and recommendations from CompassQL. For this reason, this
approach is not practical yet and more work is needed to collect enough high
quality training data. A main concern with the quality of the training data is
also whether the data is representative of the true distribution of visualizations
in the wild. Machine learning models are probabilistic models that rank examples
higher if they have seen many similar examples in their training corpus. Consequently,
if the training data is biased, the model may only recommend a single
visualization type. Understanding how the bias in training data affects neural models
is an open area of research.

\section{Next Steps in Learning Visualization Models}\label{sec:agenda}
We view the existing body of work as the first step towards moving beyond heuristics and learning visualization design from data. Multiple avenues for future work lie in designing more interpretable machine learning models, developing scalable methods for collecting different forms of training data, and designing adaptive and evolving models. 

\subsection{Tooling to Help Designers Understand Models}
Machine learning models can remove the manual effort of writing and tweaking rules by building rules on the fly. Moreover, as we are getting more data, it is easy to retrain the machine learning models rather quickly and frequently, thus improving the accuracy of the models. Despite the advantages of learning
models, potential downsides of incorporating machine learning models include an extra layer of complexity and diminished transparency~\cite{Abdul2018}, both of which can make it difficult for both developers and end users to understand how the system works. Unlike cases where the designers apply visualization design guidelines manually, incorporating machine learning models might result in losing the ability to look at the designed guidelines and their underlying
rationale. 

Going forward, designers incorporating the empirical data to learn models should provide methods to better investigate the underlying rationale and convey an understanding of how the trained models will behave in the future. Such visualization systems should communicate a variety of technical information, such as the most representative features used in training the model, the level of correlation among different features, and others. This would help designers better understand the underlying logic behind rules extracted by the model. Moreover, we envision visualization systems combining machine learning models with state-of-the-art human-computer interface techniques capable of translating models into understandable and useful explanation dialogues for end users. Such systems should explain to the end users such things as: why does the learned model recommend a specific set of visualizations? What factors does the learned model use to prune and rank visualizations? How do user interactions with the system affect the recommendations?

\subsection{Collecting Training Data}

In order to learn the design of effective visualizations from user data, access to high quality data with sufficient variety and context is critical.

Perceptual studies measuring effectiveness as defined in a controlled
setting can provide training data to learn design guidelines. However, current data from graphical perception studies are stored in different destinations all over the web (\eg different repositories, personal webpages, \etc). These inconsistencies make it harder for researchers and designers to track and access available datasets. Better organizing data collected by graphical perception studies can make deriving models easier. Going forward, one solution might be to create a single data repository where the community can share data collected from these empirical studies, thus improving the accessibility of the collected data. 

Graphical perception experiments often have specific research questions they set out to answer. These studies are typically conducted under different conditions, varying sample sizes, varying datasets, and for a diverse set of tasks. As such, they may provide useful but inherently partial and incomplete data to be used as training data for a generalizable models. Going forward, we need large-scale, focused data collection efforts to provide the desired consistency, size, context, and variation required to train generalizable models with large learning capacities. Active learning methods can guide the data collection. 

An alternative to collecting new data is to use existing corpora of visualizations on the web, such as
Tableau Public~\cite{tableaupublicmorton}, Beagle~\cite{battle2018beagle}, and the Chartmaker directory\footnote{\url{http://chartmaker.visualisingdata.com/}} or figures in papers from Viziometrics~\cite{lee2016viziometrix}. The design principles learned
from this data would reflect the reality of the kinds of visualizations scientists,
data analysts, and journalists create. However, the data may be confounded by the tools
used in a particular community, as well as network effects. 

% \todo{Cite Leilani's work, mit/harvard's massvis, Manolis' revision, Jorge's reverse engineering}

% \section{Applying Data Driven Models of Visualization Design Guidelines}
% % How can a model be applied in practice? Talk about all the cool use cases we have in mind. 

% \subsection{Evaluating Machine Learning Models}
% \bs{Jeff's comment}

\subsection{Adapting and Evolving Models}
All systems have only partial knowledge of context and user intent. For example, the analyst's goals often depend on seeing a rendered chart with real data before they realize what needs to be adjusted. As a specific example, an analyst may decide to use a log scale upon seeing that the spread of data is very large. Thus, it is crucial that recommendation systems support iterative refinement of users' intent. Moreover, individual preferences may mean that the same model is not optimal for everyone. A core concern in machine learning for visualization design should be how models can be adapted to the group or individual using it. As such, we need to develop models that take into account user feedback during visual data exploration. Ideally, the accuracy of such models should increase as users interact with the system.  

% This breaks the flow of the paper
% There is also a fundamental tension in any recommender system where increased automation usually decreases control. Automated visualization design systems such as CompassQL and Draco address this tension by letting users override any decisions that the system has made in addition to letting them iteratively refine their queries.

Even a model that adapts to users will never provide perfect recommendation, as the models are limited. Current models are restricted by the number of features they use and the data they were trained on. Models need to evolve and expand as more data becomes available and researchers find new design rules.
% One naive solution to this problem is that designers manually combine the most recent data with old data and train the model periodically.
However, this requires designers to spend a tremendous amount of time and effort to combine existing and new data since the data are collected in different formats. One possible next step towards creating adapting and evolving models is to find a common language/format and destination to collect the results of the empirical studies. Ideally, we can develop systems that incorporate the incoming data and update the model automatically.

\section{Conclusion}
In the past, visualization recommendation systems have incorporated visualization design guidelines through a manual process of curation and application. In this paper, we argue it is time to move beyond this laborious process that is hard to scale. We imagine a future in which these systems are machine learning models learned from experimental data. To achieve this future, steps must be taken in engineering robust and adaptable models, developing tools for interpretability of the created models, and consolidating data. Once achieved, however, new guidelines and data collected from graphical perception studies can be applied in practice faster and in an unbiased and reproducible manner.

\subsection*{Acknowledgements}
We thank the reviewers and Kevin Hu for their feedback.

\bibliographystyle{abbrv-doi}
\bibliography{main}

\begin{thebibliography}{10}

\bibitem{Abdul2018}
A.~Abdul, J.~Vermeulen, D.~Wang, B.~Y. Lim, and M.~Kankanhalli.
\newblock Trends and trajectories for explainable, accountable and intelligible
  systems: An hci research agenda.
\newblock CHI '18, 2018. doi: {{%
10\hspace{.1pt}\discretionary{.}{%
}{.}\hspace{.4pt}1145\discretionary{/}{%
}{/}3173574\hspace{.1pt}\discretionary{.}{%
}{.}\hspace{.4pt}3174156}}


\bibitem{aggarwal2016recommender}
C.~C. Aggarwal et~al.
\newblock {\em Recommender systems}.
\newblock Springer, 2016.

\bibitem{relationalmodels}
P.~W. Battaglia, J.~B. Hamrick, V.~Bapst, A.~Sanchez-Gonzalez, V.~Zambaldi,
  M.~Malinowski, A.~Tacchetti, D.~Raposo, A.~Santoro, R.~Faulkner, C.~Gulcehre,
  F.~Song, A.~Ballard, J.~Gilmer, G.~Dahl, A.~Vaswani, K.~Allen, C.~Nash,
  V.~Langston, C.~Dyer, N.~Heess, D.~Wierstra, P.~Kohli, M.~Botvinick,
  O.~Vinyals, Y.~Li, and R.~Pascanu.
\newblock {R}elational inductive biases, deep learning, and graph networks,
  2018.
\newblock arXiv:1806.01261v1.

\bibitem{battle2018beagle}
L.~Battle, P.~Duan, Z.~Miranda, D.~Mukusheva, R.~Chang, and M.~Stonebraker.
\newblock Beagle: Automated extraction and interpretation of visualizations
  from the web.
\newblock In {\em Proceedings of CHI}. ACM, 2018.

\bibitem{bertin1983semiology}
J.~Bertin.
\newblock Semiology of graphics: diagrams, networks, maps.
\newblock 1983.

\bibitem{cleveland1984}
W.~S. Cleveland and R.~McGill.
\newblock Graphical perception: {T}heory, experimentation, and application to
  the development of graphical methods.
\newblock {\em Journal of the American Statistical Association}, 1984.

\bibitem{perceptualKernels}
{\c{C}}.~Demiralp, M.~Bernstein, and J.~Heer.
\newblock Perceptual kernels for visualization design.
\newblock {\em IEEE Vis (Proc. InfoVis)}, 2014.

\bibitem{data2vis}
V.~Dibia and {\c{C}}.~Demiralp.
\newblock {Data2Vis}: Automatic generation of data visualizations using
  sequence to sequence recurrent neural networks.
\newblock {\em CoRR}, abs/1804.03126, 2018.

\bibitem{fechner1966elements}
G.~Fechner.
\newblock Elements of psychophysics.
\newblock 1966.

\bibitem{heer:chi09}
J.~Heer, N.~Kong, and M.~Agrawala.
\newblock Sizing the horizon: The effects of chart size and layering on the
  graphical perception of time series visualizations.
\newblock In {\em ACM Human Factors in Computing Systems (CHI)}, 2009.

\bibitem{vizdeck2012}
A.~Key, B.~Howe, D.~Perry, and C.~R. Aragon.
\newblock Vizdeck: self-organizing dashboards for visual analytics.
\newblock In {\em Proceedings of the {ACM} {SIGMOD} International Conference on
  Management of Data, {SIGMOD} 2012, Scottsdale, AZ, USA, May 20-24, 2012},
  2012. doi: {{%
10\hspace{.1pt}\discretionary{.}{%
}{.}\hspace{.4pt}1145\discretionary{/}{%
}{/}2213836\hspace{.1pt}\discretionary{.}{%
}{.}\hspace{.4pt}2213931}}


\bibitem{kim2018}
Y.~Kim and J.~Heer.
\newblock Assessing effects of task and data distribution on the effectiveness
  of visual encodings.
\newblock {\em Proc. EuroVis}, 2018.

\bibitem{Kok2005}
S.~Kok and P.~Domingos.
\newblock Learning the structure of markov logic networks.
\newblock In {\em Proceedings of the 22Nd International Conference on Machine
  Learning}, ICML '05. ACM, New York, NY, USA, 2005.

\bibitem{lee2016viziometrix}
P.~Lee, J.~West, and B.~Howe.
\newblock Viziometrix: A platform for analyzing the visual information in big
  scholarly data.
\newblock In {\em BigScholar Workshop (co-located at WWW)}, 2016.

\bibitem{deepEye:2018}
Y.~Luo, X.~Qin, N.~Tang, G.~Li, and X.~Wang.
\newblock Deepeye: Creating good data visualizations by keyword search.
\newblock In {\em Proceedings of the 2018 International Conference on
  Management of Data}, SIGMOD '18, pp. 1733--1736. ACM, New York, NY, USA,
  2018.

\bibitem{Mackinlay1986}
J.~Mackinlay.
\newblock Automating the design of graphical presentations of relational
  information.
\newblock {\em ACM Transactions on Graphics}, 1986.

\bibitem{Mackinlay2007}
J.~D. Mackinlay, P.~Hanrahan, and C.~Stolte.
\newblock {Show Me}: Automatic presentation for visual analysis.
\newblock {\em IEEE Vis (Proc. InfoVis)}, 13, 2007.

\bibitem{Mittal1998}
V.~O. Mittal, G.~Carenini, J.~D. Moore, and S.~Roth.
\newblock {Describing complex charts in natural language: A caption generation
  system}.
\newblock {\em Computational Linguistics}, 1998.

\bibitem{draco}
D.~Moritz, C.~Wang, G.~Nelson, H.~Lin, A.~Smith, B.~Howe, and J.~Heer.
\newblock Formalizing visualization design knowledge as constraints: Actionable
  and extensible models in draco.
\newblock In {\em IEEE Vis (Proc. InfoVis)}, 2018.

\bibitem{tableaupublicmorton}
K.~Morton, M.~Balazinska, D.~Grossman, R.~Kosara, and J.~Mackinlay.
\newblock Public data and visualizations: How are many eyes and tableau public
  used for collaborative analytics?
\newblock {\em SIGMOD Rec.}, 2014.

\bibitem{Quinlan1990}
J.~R. Quinlan.
\newblock Learning logical definitions from relations.
\newblock {\em Machine Learning}, 1990. doi: {{%
10\hspace{.1pt}\discretionary{.}{%
}{.}\hspace{.4pt}1007\discretionary{/}{%
}{/}BF00117105}}


\bibitem{Richardson2006}
M.~Richardson and P.~Domingos.
\newblock Markov logic networks.
\newblock {\em Machine Learning}, 62, 2006. doi: {{%
10\hspace{.1pt}\discretionary{.}{%
}{.}\hspace{.4pt}1007\discretionary{/}{%
}{/}s10994\discretionary{%
}{-}{-}006\discretionary{%
}{-}{-}5833\discretionary{%
}{-}{-}1}}


\bibitem{saket2018}
B.~Saket, A.~Endert, and {\c{C}}.~Demiralp.
\newblock Task-based effectiveness of basic visualizations.
\newblock {\em IEEE Vis (Proc. InfoVis)}, 2018.

\bibitem{Satyanarayan2017}
A.~Satyanarayan, D.~Moritz, K.~Wongsuphasawat, and J.~Heer.
\newblock {Vega-Lite}: {A} grammar of interactive graphics.
\newblock {\em IEEE Vis (Proc. InfoVis)}, 2017.

\bibitem{stevens1957psychophysical}
S.~S. Stevens.
\newblock On the psychophysical law.
\newblock {\em Psychological review}, 1957.

\bibitem{Szafir2018}
D.~A. Szafir.
\newblock Modeling color difference for visualization design.
\newblock {\em IEEE Vis (Proc. InfoVis)}, Jan 2018. doi: {{%
10\hspace{.1pt}\discretionary{.}{%
}{.}\hspace{.4pt}1109\discretionary{/}{%
}{/}TVCG\hspace{.1pt}\discretionary{.}{%
}{.}\hspace{.4pt}2017\hspace{.1pt}\discretionary{.}{%
}{.}\hspace{.4pt}2744359}}


\bibitem{talbot:infovis11}
J.~Talbot, J.~Gerth, and P.~Hanrahan.
\newblock Arc length-based aspect ratio selection.
\newblock {\em IEEE Trans. Visualization \& Comp. Graphics}, 2011.

\bibitem{Wongsuphasawat2016}
K.~Wongsuphasawat, D.~Moritz, A.~Anand, J.~Mackinlay, B.~Howe, and J.~Heer.
\newblock Voyager: Exploratory analysis via faceted browsing of visualization
  recommendations.
\newblock {\em IEEE Vis (Proc. InfoVis)}, 2016.

\end{thebibliography}

\end{document}